\documentclass[copyright,creativecommons]{eptcs}
\providecommand{\event}{LFMTP 2026}
\usepackage[dvipsnames,svgnames,table]{xcolor}
\usepackage{xspace}
\usepackage{amsmath}
\usepackage{amssymb}
\usepackage{amsthm}
\usepackage{thmtools}
\usepackage{cmll} 
\usepackage{dsfont}   
\usepackage{letltxmacro}
\usepackage{listings}
\usepackage{mleftright}
\usepackage{proof}
\usepackage{doi}
\usepackage{stmaryrd} 
\usepackage{hyperref}

\usepackage{relsize}

\colorlet{lprolog}{blue!70!black}
\colorlet{abellatop}{blue!70!green}
\colorlet{abellatac}{orange!30!black}
\colorlet{abellabad}{red!80!yellow}

\lstdefinelanguage{lprolog}{%
  alsoletter={-},
  classoffset=0,%
  morekeywords={sig,module,type,kind,pi,sigma,end},%
  keywordstyle=\color{lprolog},%
  classoffset=0,%
  otherkeywords={:-,=>,<=,\&},%
  sensitive=true,%
  morestring=[bd]",%
  morecomment=[l]\%,%
  morecomment=[n]{/*}{*/},%
}

\lstdefinelanguage{abella}[]{lprolog}{%
  alsoletter={-},
  classoffset=1,%
  morekeywords={Close,CoDefine,Define,Kind,Query,Quit,Specification,
    Set,Split,Theorem,Type,Undo,by,as,prop,true,false,forall,exists,nabla},%
  keywordstyle=\color{abellatop},%
  classoffset=2,%
  morekeywords={abbrev,apply,backchain,case,coinduction,cut,fchain,
    induction,inst,intros,monotone,on,permute,rename,left,right,
    witness,saturate,search,split,to,unabbrev,unfold,assert,with},%
  keywordstyle=\color{abellatac},%
  classoffset=3,%
  morekeywords={undo,abort,skip,clear},%
  keywordstyle=\color{abellabad}\underbar,%
  classoffset=0,%
}

\newcommand*{\SavedLstInline}{}
\LetLtxMacro\SavedLstInline\lstinline
\DeclareRobustCommand*{\lstinline}{%
  \ifmmode
    \let\SavedBGroup\bgroup
    \def\bgroup{%
      \let\bgroup\SavedBGroup
      \hbox\bgroup
    }%
  \fi
  \SavedLstInline
}
\DeclareRobustCommand\lsti[1][]{\lstinline[basicstyle=\ttfamily,keepspaces=true,#1]}

\newcommand{\RED}[1]{{\color{Crimson}\;#1\;}}
\newcommand{\BLUE}[1]{{\color{blue}\;#1\;}}
\newcommand{\blue}{\color{blue}}

\newcommand{\ie}{i.e.}

\newcommand{\Cscr}{\mathcal{C}}

\newcommand{\ra}{\rightarrow}

\newcommand{\twoseq}[2]{#1\vdash #2}
\newcommand{\Colon}{\mathrel{:\kern-.5pt:}}
\newcommand{\cutud}{cut\kern 1pt\mathord{\Updownarrow}}
\newcommand{\mcutud}{mcut\kern 1pt\mathord{\Updownarrow}}
\newcommand{\iscdot}[1]{\ifthenelse{\equal{#1}{\cdot}}}
\newcommand{\Strut}{\relax}
\newcommand{\jUnf}[4] 
  {\Strut #1\iscdot{#2}{~}{\BLUE{\mathord\Uparrow\, #2}}\mathord
   \vdash\iscdot{#3}{~}{\BLUE{#3\,\mathord\Uparrow}}\iscdot{#4}{ }{#4}}
\newcommand{\jf  }[4] 
  {\Strut #1\iscdot{#2}{~}{\RED{\Downarrow#2}}
     \mathord\vdash
     \iscdot{#3}{~}{\RED{#3\Downarrow }}
     \iscdot{#4}{ }{#4}}
\newcommand{\njf  }[4] 
  {\Strut #1\iscdot{#2}{~}{\RED{\Downarrow#2}}
     \mathord\nvdash
     \iscdot{#3}{~}{\RED{#3\Downarrow }}
     \iscdot{#4}{ }{#4}}

\newcommand{\LJE}    {\hbox{\sl LJ\kern 2pt$_\Equivs$}\xspace}
\newcommand{\LJce}   {\hbox{\sl LJ\kern 2pt$^+$}\xspace}
\newcommand{\eqLJ}   {\hbox{\sl LJ\kern 2pt$^=$}\xspace}
\newcommand{\eqLJE}  {\hbox{\sl LJ\kern 2pt$^=_\Equivs$}\xspace}

\newcommand{\Equivs }{\mathds{E}}  

\newcommand{\isemp}[1]{\ifthenelse{\isempty{#1}}{\cdot}{#1}}

\newcommand{\typeof}[2]{\hbox{\sl typeof}~#1~#2}
\newcommand{\intg}{\hbox{\sl int}}

\newcommand{\NSeq}[3]{#1 : #2 \longrightarrow #3}

\newcommand{\Judg}[2]{#1 \mathbin{\triangleright} #2}
\newcommand{\Judgc}[3]{{\color{#1} #2\;\mathbin{\triangleright}\;} #3}

\theoremstyle{remark}

\newcommand{\lP}{$\lambda$Prolog\xspace}
\newcommand{\dtl}{dependently typed $\lambda$}

\title{Proof Theory and Dependent Type Theory:\\
       Distinct Foundations for Designing Proof Assistants}
\author{Dale Miller
  \institute{Inria Saclay and LIX, Institut Polytechnique de Paris}}
\def\titlerunning{Proof Theory and Dependent Type Theory}
\def\authorrunning{D. Miller}

\begin{document}
\maketitle
\begin{abstract}
This paper examines the foundational distinctions between proof theory
and dependent type theory (DTT) in the design of interactive theorem
provers. While several implemented systems are designed using the
\dtl-calculus to represent proofs, no major proof assistant is
designed using modern structural proof theory, even though, as I will
argue here, the sequent calculus offers a compelling alternative
framework. Six specific topics are proposed where the proof-theoretic
perspective is arguably superior to the DTT perspective. These topics
include the separation of logic from proof structure, the strategic
use of non-determinism in proof reconstruction, and the avoidance of
complex typing-discipline issues such as universe levels and proof
irrelevance. The final topic---the treatment of bindings---is further
developed to demonstrate how a natural, intensional approach is
achieved through the mobility of binders. This methodology is
illustrated via the Abella theorem prover, which leverages
$\lambda$-tree syntax and the $\nabla$-quantifier to provide an
elegant environment for reasoning about the meta-theory of languages
and logics involving complex binding.
\end{abstract}

\section{Introduction}
\label{sec:intro}

In 2006, Wiedijk published a small volume \cite{wiedijk06book} that
showed the formalization of the proof of the irrationality of
$\sqrt{2}$ in the following seventeen interactive theorem provers:
\begin{quote}
ACL2, Alfa/Agda, B Method, Coq, HOL, IMPS, Isabelle/Isar, Lego,
Metamath, Mizar,\\
Minlog, Nuprl, $\Omega$mega, Otter/Ivy, PVS, PhoX, and Theorema.
\end{quote}
These systems vary significantly in their underlying logical
foundations. Some are based on first-order logic, while others accept
higher-order quantification. Employing a range of proof structures,
some of these systems use Frege-style proofs (involving axioms and a
small set of inference rules) because these can support simple
proof-checking kernels. Others take a more sophisticated approach to
proof structures and employ either a natural deduction style \`a~la
Gentzen and Prawitz, or the \dtl-calculus directly. The role of
\emph{types} in these systems varies greatly. Most of these systems
allow at least multi-sorted, first-order quantification. Others
implement Church's Simple Theory of Types and allow quantification
over function and predicate types. In addition, certain predicates
that have well-understood reasoning principles can often be recast as
types, with the usual consideration of whether type checking and/or
type inference are decidable for those predicates cum types. Some of
these systems (and the more recent Lean system~\cite{moura15cade}),
are based on \emph{dependent type theory} (DTT), whose origin is often
traced back to De Bruijn~\cite{debruijn80} and Martin-L\"{o}f
\cite{martinlof84}.

In this paper, I argue that modern aspects of proof theory can offer
an alternative perspective to the design narrative behind the use of
\dtl-calculus in proof assistants.  I will assume here that the reader
is familiar with the latter formalism but is not familiar with the
many advances that have occurred within the general topic of
structural proof theory.  I will survey that topic in the next
section.

\section{Structural proof theory}

Structural proof theory was started by Gentzen~\cite{gentzen35} with
his invention of natural deduction and the sequent calculus.  In
particular, Gentzen described his intentions about natural deduction:
\begin{quote}
    I intended first to set up a formal system which comes as close as
    possible to actual reasoning.  The result is a \emph{`calculus of
    natural deduction'}.
\end{quote}
Thus, it is no surprise that this calculus of natural deduction
has had important influences on various proof assistants.

Early in the introduction of \cite{gentzen35}, Gentzen justifies the
introduction of the sequent calculus as a technical vehicle for
stating and proving the \emph{Hauptsatz} for \emph{both}
intuitionistic and classical logic.  In the general area of
computational logic, sequent calculus is often seen as a technical
device for tracing the search for an actual proof.  In fact, there is
the old chestnut: ``Natural deductions are proofs while sequent
calculus proofs are the computation of a proof'' (see, for example,
\cite[Section 5.4]{girard89book}).

As we shall see in this paper, the sequent calculus can be viewed, not
only as a useful technical device, but as a central framework for
developing and justifying possible proof assistant designs.  It is
also possible to list several successful deployments of the sequent
calculus in domains other than proof system design.
\begin{itemize}
  \item Gentzen used the sequent calculus to prove the consistency of
    Peano arithmetic~\cite{gentzen36ent}.
  \item Various researchers, for example
    \cite{ketonen22book,negri11book}, have used the sequent calculus
    to provide independence proofs in certain axioms with respect to
    other axioms within mathematical theories.
  \item Often, the most successful proof system for a modern logic, such as
    modal and linear logics, is given using the sequent calculus.
  \item Various logic programming languages have been developed using
    the sequent calculus in classical, intuitionistic, and linear
    logics~\cite{miller25ptlp}.
  \item The design of at least one model checker has been built on a
    sequent calculus proof theory that accounts for fixed
    points~\cite{baelde07cade,heath19jar}.
\end{itemize}

Of the many recent developments in the theory of the sequent calculus
that might not be familiar to those working on designing and
implementing proof systems, we can list the notions of \emph{polarity}
and \emph{focusing} (both inspired by the sequent calculus of linear
logic) and the notion of the \emph{mobility of binders}.  These topics
will be expanded on in the following section.

\section{A proof theorist's view of dependent type theory}

In this section, I identify six topics where a proof-theoretic
perspective offers distinct advantages over that of DTT. Naturally,
one could conversely argue that DTT provides a superior framework in
other areas; its strengths include mature implementations, widespread
adoption, robust support for automated inference, and the ability to
provide concise specifications in a variety of domains. My objective
here is to articulate the proof-theoretic perspective, as it is seldom
represented in the literature. Furthermore, by focusing on
foundational issues, I relegate the current state of software
implementation to a secondary concern.

\subsection{Regarding the structure of proof}

A given DTT usually settles two questions simultaneously: 
\begin{enumerate}
  \item Which logic is being formalized?  This is typically
    intuitionistic logic.
  \item What is a proof?  These are typically natural deduction proofs
    encoded as \dtl-terms.
\end{enumerate}
A proof theorist separates these questions, thereby allowing for a
possibly large variety of proof systems for a fixed logic.  For
example, when fixing the logic to be intuitionistic, proofs can be
formalized using natural deduction, sequent calculus, and
tableaux~\cite{fitting83}, 
and inference rules can be applied in a forward chaining as well as a
backward chaining fashion.  When fixing the logic to be classical
logic, many more proof structures are possible, additionally including
Herbrand disjunctions, resolution refutations, natural deduction with
restart~\cite{gabbay84}, etc.  If the logic is fixed to be linear,
then all the previous proof structures are possible, including various
forms of proof nets.

Of course, once one has a solid implementation of, say, natural
deduction proofs as \dtl-terms, one has a highly expressive
computational setting capable of encoding a rich collection of other
proof systems, as witnessed by the many proof systems that have been
encoded into Dedukti~\cite{assaf23arxiv}.  Still, there is something
to be said for treating these other proof systems as structures in
their own right rather than merely as encodings into some other proof
structure. 

\subsection{Problems with the proof-as-$\lambda$-term approach}

Another frequently claim is that the $\lambda$-calculus is a
canonical computational model for sequential computation.  While I am
not arguing against that particular sentiment, it is worth noting that
many things about the (untyped) $\lambda$-calculus are far from
canonical.  For example, while $\lambda$-reduction is a confluent
operation on $\lambda$-terms, the paths to (weak) normal forms are
highly non-deterministic, with different computational platforms
adopting different reduction strategies (call-by-value, call-by-name,
call-by-need, etc), and these can affect correctness (e.g.,
call-by-value may not terminate with a normal form while call-by-name
does) as well as efficiency.

Dependently typed $\lambda$-terms also introduce many complications.
For example, there are issues around proof irrelevance (too many
subproofs kept), implicit arguments (too inconvenient to supply all
arguments), and universe levels (needed to organize rich typing
structures).  Approaches to encoding logic that omit such a rich and
complex typing discipline do not need to address these issues, at
least not in a foundational setting.  Also, proofs-as-$\lambda$-terms
can be highly redundant (leading to, for example, explicitly and
implicitly typed versions of LF \cite{pientka13jfp,reed08lfm}) and
they do not include any explicit structure-sharing mechanisms.  The
sequent calculus, on the other hand, provides a simple and natural
explicit mechanism~\cite{miller23csl}.

\subsection{The mutual development of dependent type theory and proof theory}

Many central topics in the study of proof theory and DTT were
investigated around the same time and influenced each other. 
\begin{enumerate}
\item The notion of normalizing proofs was first developed in the
  context of sequent calculus (via cut-elimination~\cite{gentzen35})
  and natural deduction (via
  normalization~\cite{plato08bsl,prawitz65}).  These normalization
  processes (including the elimination of non-atomic initial rules)
  yield the familiar notions of $\beta$ and $\eta$-rules.  Of course,
  these normalization processes were also studied in the untyped and
  simply typed $\lambda$-calculus by Church also in the 1930s and
  1940s~\cite{church32am,church41}.  The introduction of these concepts
  into \dtl-calculus waited until de
  Bruijn~\cite{debruijn80} and Martin-L\"of~\cite{martinlof82}
  introduced their typed calculi.

\item Linear logic \cite{girard87tcs} appeared first as a development
  in proof theory and later moved to \dtl-calculus shortly after (see,
  for example, \cite{cervesato96lics,hofmann03ic}).

\item The notion of \dtl-terms in \emph{canonical form} (as defined
  for LF in~\cite{harper93jacm}) is closely related to the proof
  theoretic concept of \emph{uniform proofs} (as shown in
  \cite{felty91lf,felty90cade}).  Eventually, the linear logic notions
  of \emph{polarization} and \emph{focusing}~\cite{andreoli92jlc} were
  generalized within the proof theory setting to classical and
  intuitionistic logics (see, for example,
  \cite{danos93wll,dyckhoff06cie,liang09tcs}).  Focused proofs are now
  generally understood to yield useful notions of canonical normal
  forms of proofs~\cite{chaudhuri08tcs,miller25fscd}.
\end{enumerate}

\subsection{The dependent type theory approach to classical logic}

Gentzen \cite{gentzen35} abandoned natural deduction since he could
not see how that style of proof could be used uniformly to yield the
Hauptsatz for both classical and intuitionistic logic.  His solution
for a more uniform approach to the proof theory of both logics was a
multiple-conclusion sequent calculus.  Unfortunately, Gentzen's better
approach is generally ruled out in DTT, at least in the type
theories in use in proof assistants today.  The gap between treating
classical logic proofs as \dtl-terms plus axioms, such as the excluded
middle, and viewing classical logic proofs, such as say Herbrand
disjunctions (expansion trees) or resolution refutation, seems huge.
Of course, elaborate techniques have been developed to bridge this
gap, at least, in principle: see, for example,
\cite{ebner18paar,komel25arxiv}. 

\subsection{Non-determinism provides a valuable resource}

The trusted proof-checking kernels of proof assistants based on
\dtl-calculus are generally functional programs that expect the proof
structures they check to contain enough information to make checking
completely deterministic.  It is well known that non-determinism can
be a valuable resource in designing and verifying certificates: just
consider the P and NP complexity classes.  Allowing
non-determinism in the design of both proof certificates and proof
checkers allows for important trade-offs between proof certificate
size and checking time.

Having a kernel that allows non-determinism (via backtracking
search) is certainly possible (as in logic-programming-based
kernels).  Example: Consider a proof certificate for the claim that
the sequent $\Gamma\vdash A$ is an instant of the \emph{initial} rule,
which can only be the case when $A$ is a member of $\Gamma$.  On one
hand, the certificate could be constant in size while the proof
checking kernel would then need to search for $A$ in $\Gamma$:
something that could be accomplished with a non-deterministic search
(made deterministic in actual deterministic kernel using search).  On
the other hand, the certificate could provide the exact name/label of
the assumption in $\Gamma$ that is also labeled with $A$.  In this
case, the certificate is no longer constant-sized, but its checking
would be a more direct, deterministic process.  Having this trade-off
between proof certificate size and non-deterministic proof
reconstruction seems desirable.  A kernel for such checking is a
rather direct combination of some form of (higher-order) unification
and backtracking search.  While incorporating such features into the
trusted code base of a kernel will increase the complexity of such
kernels, these features have been well-developed and exploited in
proof systems written in the $\lambda$Prolog programming
language~\cite{miller12proghol} and the Isabelle prover
\cite{paulson86jlp} since the late 1980s.

\subsection{Treatment of bindings}

In most modern proof assistant based on DTT or first-order logic, no
canonical treatment for bindings has appeared.  Instead, a wide range
of technical approaches to encoding bindings have been deployed and
studied.  In general, the specifications of bindings in syntax are
still considered problematic in most conventional systems, given the
existence of challenge problems (POPLMark~\cite{aydemir05tphols},
POPLMark reloaded~\cite{abel19jfp}), case studies, benchmarks, and
surveys.  The Twelf~\cite{pfenning99cade} and Beluga
\cite{pientka10ijcar} systems are exceptions since they offer a
computing environment where computed values can be \dtl-terms.
However, of the proof assistants lists in Section~\ref{sec:intro},
only Isabelle and Minlog offer direct support for bindings in data
structures and the associated notion of (higher-order) unification on
such structures.

The sequent calculus provides an elegant and powerful setting to
specify and compute with syntactic objects containing bindings,
something that, to date, has not been matched in the setting of
\dtl-calculi.  The approach available in the sequent calculus is
outlined in more detail in the following section.

\section{Sequents and binders}

By making a slight embellishment of Gentzen's formulation of sequents,
we write sequents here as $\Sigma:\twoseq{\Gamma}{C}$, where the
collection of variables in $\Sigma$ is interpreted as being bound over
$\twoseq{\Gamma}{C}$.  The variables in $\Sigma$ correspond to
Gentzen's notion of \emph{eigenvariables}.

To illustrate how binder can be used with sequent calculus proof
rules, consider specifying the predicate 
\hbox{\sl typeof}, which relates an encoding of untyped
$\lambda$-terms (given by the function $\lceil\cdot\rceil$) with an
encoding of simple types (using $\ra$ for the function type
constructor).  Here, we assume as is customary in many encodings of
$\lambda$-terms (see, \cite{miller12proghol}, for example) that binders
in the untyped $\lambda$-terms are mapped to bindings in the result of
the encoding.

When reading inference rules from conclusion to premises (as one does
to understand the impact of an inference rule during the search for a
proof), binders can be \emph{instantiated}, as in the following
instance of the left-introduction for $\forall$.
\[
  \infer[\forall L]
        {\twoseq{\Sigma:\Gamma,\forall \alpha(\typeof{c}{(\alpha\ra\alpha)})}{C}}
        {\twoseq{\Sigma:\Gamma,\typeof{c}{(\intg\ra\intg)}}{C}}
\]
Here, the bound variable $\alpha$ is instantiated with the term
$\intg$, a term denoting the integer type.
Binders can also \emph{move}, as illustrated by the following
inference rules.
\[
\infer
  {\twoseq{\Sigma:\Gamma}{\typeof{\lceil{\blue\bf \lambda x}.B\rceil}
                              {(\alpha\ra\beta)}}}
  {\infer[\forall R,\mathord{\supset}R]
    {\twoseq{\Sigma:\Gamma}{{\blue\bf\forall x}(\typeof{x}{\alpha}\supset
                                             \typeof{\lceil B \rceil}{\beta})}}
    {\twoseq{\Sigma,{\blue\bf x}:\Gamma,\typeof{x}{\alpha}}
                {\typeof{\lceil B \rceil}{\beta}}}}
\]
Here, the lower inference rule is justified by some external
definition of the predicate {\sl typeof} (a familiar specification
technique for defining typing dating back to the early days of \lP and
LF~\cite{felty90cade,harper93jacm}).  Here, binders move from
\emph{term-level} ($\lambda x$) to \emph{formula-level} ($\forall x$)
to \emph{proof-level} (eigenvariable): that is, this bound variable
never becomes free.  Thus, \emph{binder mobility} is one way to
formally realize A. Perlis's Epigram 47: ``There is no such thing as a
free variable.'' \cite{perlis82sigplan}.  A consequence of this
approach to binders is that the technique for implementing binders
(e.g., named variables, nameless dummies, etc) does not need to be
revealed to the person writing the logical specification of, in this
case, \textsl{typeof}: instead, binders never stop being bound.

Another aspect of binders in the proof-theory setting is that they do
not need to represent function spaces: they can be viewed, in the
appropriate logical setting, as abstractions over syntax.
Consider, for example, the formula
\[
  \forall w.\ \lambda x.x \not= \lambda x.w,
\]
where $w$ and $x$ are given the same primitive type.
If one views the $\lambda$-binding as specifying a function (\ie,
extensionally), then this formula is not a theorem since the identity
and a constant-valued function can coincide on singleton domains.  If
one views the $\lambda$-binder as simply part of the syntax (up to
$\alpha$-conversion, of course), then this formula should be a theorem
since no (capture avoiding substitution) instance of $\lambda x.w$ can
be equal to $\lambda x.x$. 

This approach to binders in which syntactic binders are movable is
called the $\lambda$\emph{-tree syntax} approach~\cite{miller19jar},
and it is available in the \lP programming
language~\cite{miller12proghol}, the LF logical
framework~\cite{harper93jacm}, and the Rocq plugin for
Elpi~\cite{tassi26hdr}.

The sequent calculus can support at least one additional binding site
other than the prefixed $\Sigma$ binder.  In particular, we add a
binding context to each occurrence of a formula in a sequent.  Thus,
instead of the sequent $\Sigma\colon\twoseq{B_1,\dots,B_n}{B_0}$, we
have the more general
\[
\Sigma\colon\twoseq
   {\Judgc{red}{\sigma_1}{B_1}, \dots, \Judgc{red}{\sigma_n}{B_n}}
   {\Judgc{red}{\sigma_0}{B_0}}.
\]
Here, $\sigma_i$ is a list of distinct variables scoped over $B_i$.
The expression $\Judg{\sigma_i}{B_i}$ is called a
\emph{generic judgment}.  In order to exploit these proof-level
binding sites, a new formula-level binder is needed: this is the role
of the $\nabla$-quantifier, for which the left and right introduction
rules are given as
\[
\infer[\qquad\hbox{and}\qquad]
  {\NSeq{\Sigma}{\Judgc{red}{\sigma}{\nabla_\tau {\blue x}.B},\Gamma}{\Cscr}}
  {\NSeq{\Sigma}{\Judgc{red}{(\sigma, {\blue x}:\tau)}{B},\Gamma}{\Cscr}}
\infer[.]
      {\NSeq{\Sigma}{\Gamma}{\Judgc{red}{\sigma}{\nabla_\tau {\blue x}.B}}}
      {\NSeq{\Sigma}{\Gamma}{\Judgc{red}{(\sigma, {\blue x}:\tau)}{B}}}
\]
Because of the similarity of the left and right introduction rules, it
follows easily that $\nabla$ is self-dual: that is, both the sequents
$\neg\nabla x. B x\vdash\nabla x.\neg B x$ and $\nabla x.\neg B
x\vdash\neg\nabla x.B x$ are provable.  Furthermore, $\nabla$ around
an equality judgment can simply reduce to an equality judgment around
$\lambda$-terms: that is, $\nabla x. (t = s)$ is provable exactly when
$\lambda x. t = \lambda x. s$ is provable.  
Thus, the formula $\forall w.\ \lambda
x.x \not= \lambda x.w$, turns out to be logically equivalent to
$\forall w.\neg (\lambda x.x = \lambda x.w)$, which can be proved in a
suitable system (such as Abella) since the (higher-order) unification
problem $(\lambda 
x.x = \lambda x.w)$ has no unifiers.
For details on how
$\nabla$-quantification interacts with the propositional constants,
the other quantifiers, and induction and coinduction, see
\cite{gacek11ic,miller05tocl}.

The Abella proof assistant~\cite{baelde14jfr,gacek08ijcar} was
developed on proof theoretic principles and includes support for
$\lambda$-tree syntax and the $\nabla$-quantifier.

\section{The Abella proof assistant}

During the search for a proof in Abella, one works with collections of
\emph{goals}, which are displayed as 
\begin{lstlisting}
  Variables: x1 ... xm
  H1 : A1
  ...
  Hn : An
  ============================
   C
\end{lstlisting}
where $x_1, \ldots, x_m$ are universally quantified variables,
\lsti{H1}, \dots, \lsti{Hn} are \emph{hypothesis labels} that are each
associated with a unique \emph{hypothesis formula} drawn from
$A_1,\ldots,A_n$ and $C$ is a formula called the \emph{conclusion} of
the goal.  The variables and hypotheses comprise the \emph{context} of
the goal.  Not only does Abella view such a goal as a sequent, in this
case being,
\[
  x_1\colon\tau_1, \ldots, x_m\colon\tau_m \Colon A_1,\ldots,A_n\vdash C,
\]
but its tactics for advancing the search for a proof are understood
using sequent calculus inference
rules~\cite{chaudhuri25tableaux,chaudhuri18lfmtp}.   Abella implements
the $\nabla$-quantifier and the structural principles behind the
logical equivalences $(\nabla x\nabla y.B)\equiv(\nabla y\nabla x.B)$
and $(\nabla x.C)\equiv C$, provided $x$ is not free in $C$.  As a
result of these equivalences, it is possible to treat the local
binding signatures using an equivalent mechanism that relies on
nominal constants~\cite{gacek11ic} (as a result, the local signatures
are not displayed explicitly in the Abella goal structure above).

Abella is well-suited for reasoning about the meta-theory of languages
and logics, especially those involving binding.  For example, various
aspects of the meta-theory of the $\lambda$-calculus have had
successful Abella specifications: see, for example, a formulation of
the $\lambda$-cube~\cite{accattoli12cpp}, a mechanical formulation of
higher-ranked polymorphic type inference~\cite{zhao18itp}, and a
formalization of parts of Barendregt's theory of the
lambda-calculus~\cite{lancelot25itp}.  Specifications of the
operational semantics of the $\pi$-calculus and its meta-theory can
make significant use of Abella's support for binder mobility.
For example, the difference between open and closed
bisimulation for the $\pi$-calculus is a simple matter of switching
between using intuitionistic logic and classical
logic~\cite{miller05tocl}.  Similarly, the presence of the three
quantifiers $\forall$, $\exists$, and $\nabla$ provides an easy and
elegant way to capture the large collection of modal operators that
have been proposed to capture an array of properties of
$\pi$-calculus~\cite{milner93tcs,tiu10tocl}.  Additionally, rather
direct treatment of bisimulation-up-to techniques for the
$\pi$-calculus can also be captured~\cite{chaudhuri15cpp}.  The
website for Abella, \url{https://abella-prover.org/}, contains several
other example formulations using Abella.

\section{Conclusion}

In this paper, I have argued that modern structural proof theory,
rooted in the sequent calculus, offers a robust and compelling
foundational alternative to basing interactive theorem provers on the
``proofs-as-terms'' paradigm of DTT.  I have offered six
explicit topics for which the proof theory approach can be argued to
provide a possibly deeper and more flexible design than that typically
offered by \dtl-calculus.  The last of these topics---the treatment of
binding in syntactic expressions---is perhaps most significant.  In
this setting, the proof-theoretic approach provides a natural
treatment of bindings via binder mobility.  By treating binders as
movable abstractions over syntax rather than function spaces, we
achieve an elegant framework for reasoning about the meta-theory of
complex languages.  Many of these proof-theoretic notions have been
built into the Abella theorem prover: in particular, this prover makes
available the $\lambda$-tree syntax approach to bindings and the
$\nabla$-quantifier.

\medskip\noindent{\bf Acknowledgments}\quad I thank the reviewers for
their helpful comments and perspectives on an earlier draft of this
paper.

\bibliography{extract}

\end{document}